\documentclass{ckm}                 % twocolumn proceedings style
\def\KP{K^+}

\def\KS{K^0_S}
\def\KL{K^0_L}
\def\PP{\pi^+}
\def\PM{\pi^-}
\def\PZ{\pi^0}
\def\Xs{X_s}
\def\Bbar{\overline{B}{}}

\def\btogamma{b\to\gamma}
\def\btosll{b\to s\ell^+\ell^-}
\def\btosgamma{b\to s\gamma}
\def\btodgamma{b\to d\gamma}
\def\BtoXsgamma{B\to X_s\gamma}
\def\BtoXsll{B\to X_s\ell^+\ell^-}
\def\Btorhogamma{B\to\rho\gamma}
\def\BtoKll{B\to K\ell^+\ell^-}
\def\BtoKstarll{B\to K^*\ell^+\ell^-}
\def\BtoKstarG{B\to K^*\gamma}
\def\BtoKstarZG{B^0\to K^{*0}\gamma}
\def\BtoKstarPG{B^+\to K^{*+}\gamma}

\def\Br{{\cal B}}
\def\Mbc{M_{\rm bc}}
\def\DeltaE{\Delta E}
\def\Vtd{V_{td}}
\def\Vts{V_{ts}}
\def\Vub{V_{ub}}
\def\Acp{A_{CP}}
\def\MXs{M(\Xs)}
\def\Mee{M(e^+e^-)}
\def\Mll{M(\ell^+\ell^-)}

\def\MeV{\mbox{~MeV}}
\def\MeVc{\MeV/c}
\def\MeVcc{\MeVc^2}
\def\GeV{\mbox{~GeV}}
\def\GeVc{\GeV/c}
\def\GeVcc{\GeVc^2}
\def\fbinv{~\mbox{fb}^{-1}}

\usepackage{txfonts}

\confname{Workshop on the CKM Unitarity Triangle, IPPP Durham, April 2003}

\title{Belle Results on $\btosll$ and $\btogamma$}

\author{M. Nakao (for the Belle Collaboration)}
\address{KEK, High Energy Accelerator Research Organization, Tsukuba}

\begin{document}

                    %%%%%%%%%%%%%%
                    %  Abstract  %
                    %%%%%%%%%%%%%%

\begin{abstract}
We discuss the latest results and future prospects of the Belle
experiment on the electroweak and radiative processes $\btosll$,
$\btosgamma$ and $\btodgamma$.  In particular, the first measurement of
the inclusive rate for $\BtoXsll$ provides a new constraint on the
physics beyond the Standard Model.  The yet to be measured decay
$\Btorhogamma$ is expected to provide a new constraint on the CKM matrix
element $|\Vtd|$ in the near future.
\end{abstract}

\maketitle

                    %%%%%%%%%%%%%%%%%%%%%%
                    \section{Introduction}
                    %%%%%%%%%%%%%%%%%%%%%%

Radiative $B$ meson decays through the $\btosgamma$ process have been a
powerful tool to constrain physics beyond the Standard Model (SM).  The
radiated photon serves as a probe to study short distance loop
diagrams through a comparison of the decay rate with theory calculations
and through a search for direct CP asymmetry.  In addition, the
radiated photon also probes the kinematical property of $B$ decays
through the photon energy spectrum, which is useful to constrain the
lepton spectrum in $|\Vub|$ measurements.

A similar process, $\btosll$ $(\ell=e,\mu)$, acts as an additional probe
for new physics, since the existence of the $Z$-boson radiation and
$W$-boson box diagrams may enhance the effects from new physics.  The
rate is about two orders of magnitude smaller than $\btosgamma$, but the
signal
with two energetic leptons is very clean.  The lepton pair in the final
state provides two additional observables that can be used to identify
new physics effects: the differential branching fraction and the
lepton forward-backward asymmetry, as functions of the dilepton mass.
Recently Belle observed the first $\BtoKll$ events, and successfully
measured the inclusive branching fraction for $\BtoXsll$.

Another process, $\btodgamma$, is also suppressed by two orders of
magnitude with respect to $\btosgamma$, naively due to the CKM factor
$|\Vtd/\Vts|^2$.  Although the rate is similar to $\btosll$, the signal
is less clean due to huge backgrounds of $\btosgamma$ and energetic
$\PZ\to\gamma\gamma$ from the continuum $q\overline{q}$ ($q=u,d,s,c$)
production.  Assuming that the non-SM contributions to $\btosgamma$ and
$\btodgamma$ are small, we can constrain the value of $|\Vtd/\Vts|^2$
from a $\btodgamma$ measurement.  It is very important to compare the
$|\Vtd/\Vts|$ value from $\btodgamma$ with those from the
$B^0_{d,s}$-$\Bbar^0_{d,s}$ mixing, since the latter involves lattice
QCD calculations and may be affected by new physics that appear only in
the mixing diagram.

In this report, we summarize the latest Belle results on the electroweak
and radiative $B$ decays, and discuss the future prospects.  We
anticipate that a large dataset of $500\fbinv$ will be ready within the
next two to three years of KEKB/Belle operation.

                    %%%%%%%%%%%%%%%%%%%%%%%%%%%%%
                    \section{$\btosll$ processes}
                    %%%%%%%%%%%%%%%%%%%%%%%%%%%%%

After the first observation of the decay $\BtoKll$
\cite{bib:belle-kll-prl} by Belle with a $29\fbinv$ dataset, we extended
our study to perform an inclusive measurement of $\BtoXsll$, where $\Xs$
is the hadron recoil system in the $\btosll$ process
\cite{bib:belle-xsll-prl}.  We adopt a pseudo-reconstruction technique,
in which the $X_s$ system is reconstructed as a sum of one kaon ($\KP$
or $\KS$) and zero to four pions (of which up to one $\PZ$ is allowed).
Assuming the decay rate to the state with a $\KL$ is the same as the
rate with a $\KS$, this set of combinations covers $(82\pm2)\%$ of the
signal.  In addition, we do not use about 5\% of the events with
$\MXs>2.1\GeVcc$ where the background rate is large.  The $X_s$ system
is combined with a pair of electrons $(p>0.5\GeVc)$ or muons
$(p>1.0\GeVc)$ to form a beam-energy constrained mass ($\Mbc$) to
identify the signal.  The electron pairs with $\Mee<0.2\GeVcc$ are
rejected to avoid the pole at $q^2=0$ and to reduce backgrounds from
photon conversions and $\PZ\to e^+e^-\gamma$ decays.

The major background source is semileptonic $B$
decays, in which two leptons are either from two $B$ mesons or from the
subsequent charm decays.  The continuum background contribution is
smaller than semileptonic $B$ decays.  These backgrounds do not produce
signal peaks, and can be separated from the signal by fitting $\Mbc$.  In
addition, there are peaking backgrounds from $B\to X_s\PP\PM$ (mostly
from $B\to D^{(*)}n\pi$), $B\to X_s J/\psi$ and $B\to X_s \psi'$
($J/\psi,\psi'\to\ell^+\ell^-$).  The huge $B\to X_s\PP\PM$ background
becomes a contribution of $2.7\pm0.2$ events under the signal peak after
the lepton identification requirements.  The dilepton mass range around
the $J/\psi$ and $\psi'$ mass peaks are vetoed.  The peaking backgrounds
are useful control samples to evaluate various systematic errors in the
analysis.

\begin{figure} % -------------------------------------------------------
\hbox to\hsize{\hss\includegraphics[width=\hsize]{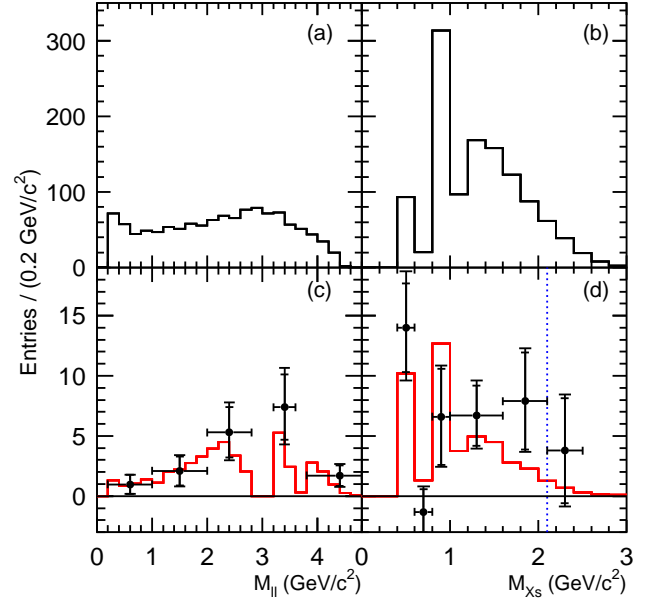}\hss}
\caption{Dilepton and recoil mass spectrum in the $\BtoXsll$.  Upper
plots shows a SM model, and lower plots show the measurements (data
points) compared with the efficiency corrected predictions (histograms).}
\label{fig:xsll}

\end{figure}

Using a data sample of $60\fbinv$, we observed about 60 events in the
$\Mbc$ distribution with an efficiency of $3.7\%$ and a statistical
significance of 5.4.  We obtained the inclusive branching fraction
\begin{equation}
\begin{array}{l}
{\displaystyle
 \Br(\BtoXsll)=(6.1\pm1.4\;^{+1.4}_{-1.1})\times10^{-6}}\\
\end{array}
\end{equation}
with a kinematical cut of $\Mll>0.2\GeVcc$; for all the other cuts, the
branching fraction is extrapolated to the entire phase space.  By
subdividing the sample into bins, we measure the $\Mll$ and $\MXs$
distributions as shown in Fig.~\ref{fig:xsll} with a comparison to the
SM expectations.  The measured inclusive branching fraction can be
compared with the SM prediction of $(4.2\pm0.7)\times10^{-6}$.  The
result is in agreement with the SM; however, the measurement error is
still too large to be conclusive.

We also performed a separate analysis for the exclusive $B\to
K^{(*)}\ell^+\ell^-$ modes with the same $60\fbinv$ dataset
\cite{bib:belle-kll-ichep02}.  We increased the precision of the $\BtoKll$ result,
but no significant signal was observed for $\BtoKstarll$ for which we
quote a 90\% confidence level upper limit.  The updated
results are
\begin{equation}
\begin{array}{l}
{\displaystyle
 \Br(\BtoKll)=(5.8\,^{+1.7}_{-1.5}\,\pm0.6)\times10^{-7}}\\
{\displaystyle
 \Br(\BtoKstarll)<14\times10^{-7}.}
\end{array}
\end{equation}

The number of events in the inclusive measurement indicates that the
first measurement of the forward-backward asymmetry will be feasible in
near future using the pseudo-reconstruction samples, even after
excluding the $\BtoKll$ contribution which does not produce an
asymmetry.  The exclusive $\BtoKstarll$ mode is also expected to be
observed soon, and a forward-backward asymmetry measurement using the
exclusive sample will be complementary.

 % ---------------------------------------------------------

                    %%%%%%%%%%%%%%%%%%%%%%%%%%%%%%%%
                    \section{$\btosgamma$ processes}
                    %%%%%%%%%%%%%%%%%%%%%%%%%%%%%%%%

\begin{figure} % -------------------------------------------------------
\hbox to\hsize{\hss\includegraphics[width=\hsize]{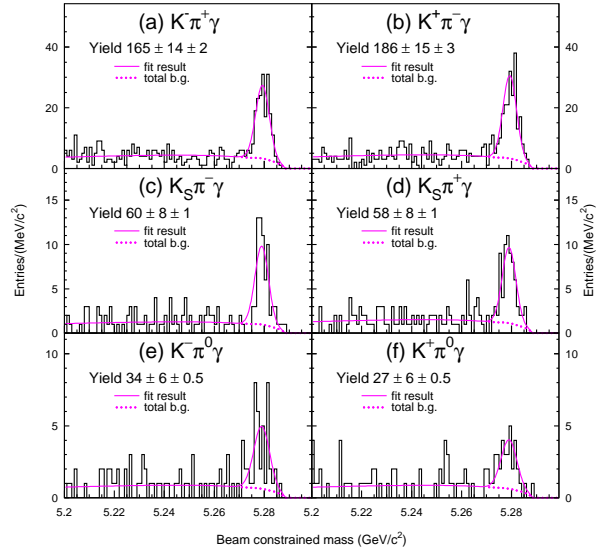}\hss}
\caption{Beam constrained mass spectrum for $\BtoKstarG$, separately
         shown for charge conjugated samples.}
\label{fig:kstargam-acp}
\end{figure}

The exclusive radiative decay $\BtoKstarG$ provides one of the most
precise measurements among $B$ meson rare decays.  Signal is clearly
seen as shown in Fig.~\ref{fig:kstargam-acp} separately for each
$K^*$ final state and charge.  We measured the $\BtoKstarG$ branching
fractions and CP asymmetry using a $60\fbinv$ dataset
\cite{bib:belle-kstargam-ichep02}
\begin{equation}
\begin{array}{l}
{\displaystyle
  \Br(\BtoKstarZG)=(39.1\,\pm2.3\,\pm2.5)\times10^{-6}}\\
{\displaystyle
  \Br(\BtoKstarPG)=(42.1\,\pm3.5\,\pm3.1)\times10^{-6}}\\
{\displaystyle
  \Acp(\BtoKstarG)=(-2.2\,\pm4.8\,\pm1.7)\times10^{-2}}.
\end{array}
\end{equation}
The branching fraction may be compared with the SM prediction, for
example, $(7\pm2)\times10^{-5}$ \cite{bib:kstargam-bosch-buchalla}.
The predicted branching fraction is higher than the measured value, but
one cannot consider this seriously due to large model-dependent
form factor uncertainties in the prediction.  We also note that the
branching fraction for the neutral and charged decays are about the same 
size, and are not yet sensitive to the isospin asymmetry proposed in
ref.~\cite{bib:kagan-neubert-2001}. 

 % ---------------------------------------------------------

On the other hand, CP asymmetry is a theoretically clean handle to
search for new physics.  A non-zero CP asymmetry is a clear indication
of new physics, since the SM predicts a very small CP asymmetry of less
than 1\%.  The measured asymmetry is already as small as 5\%, and will be
further improved by adding more data.  With an anticipated $500\fbinv$
data set, the error will be about 2\% including the systematic error,
and the total error will be still dominated by the statistics.

\begin{figure} % -------------------------------------------------------
\hbox to\hsize{%
  \hss\includegraphics[width=0.5\hsize]{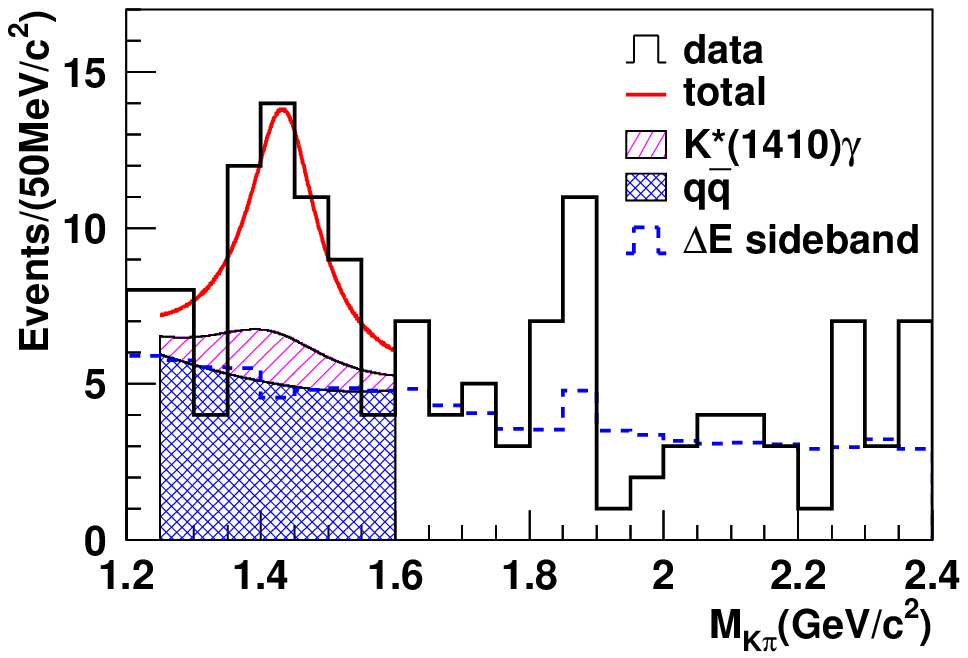}%
  \hss\includegraphics[width=0.5\hsize]{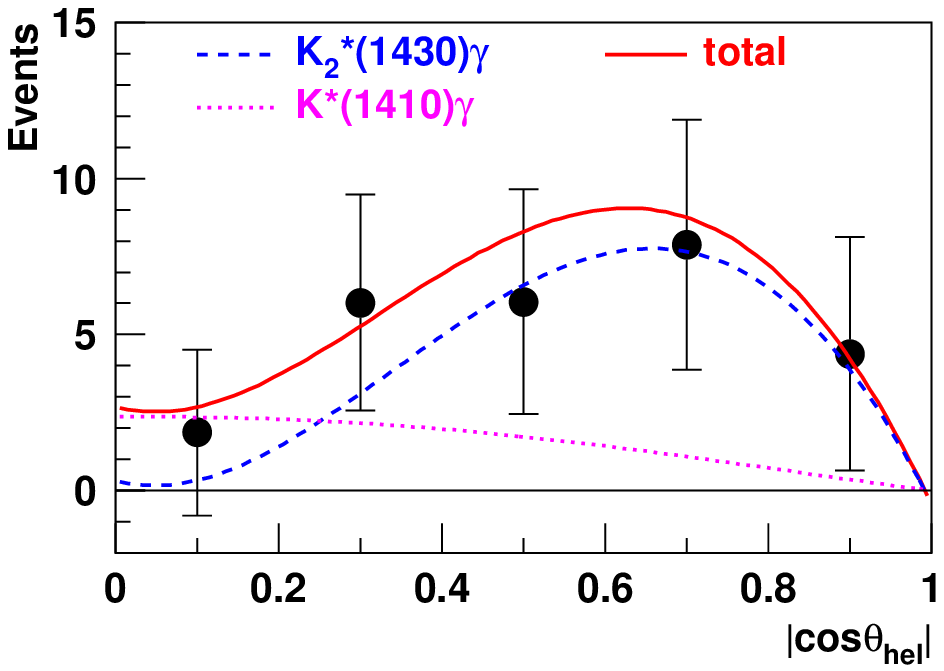}\hss}
\hbox to\hsize{%
  \hss\includegraphics[width=0.5\hsize]{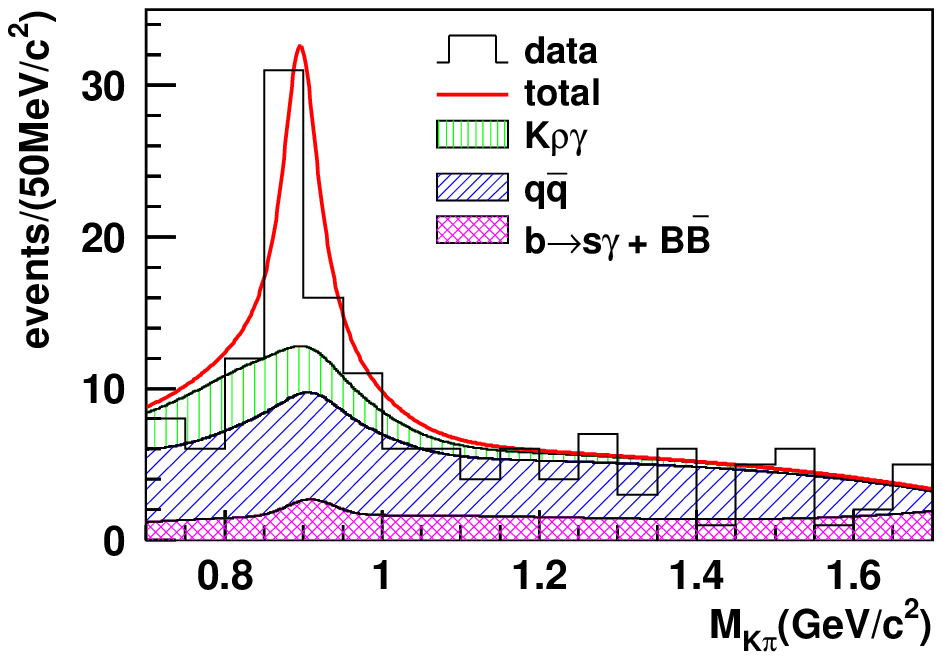}%
  \hss\includegraphics[width=0.5\hsize]{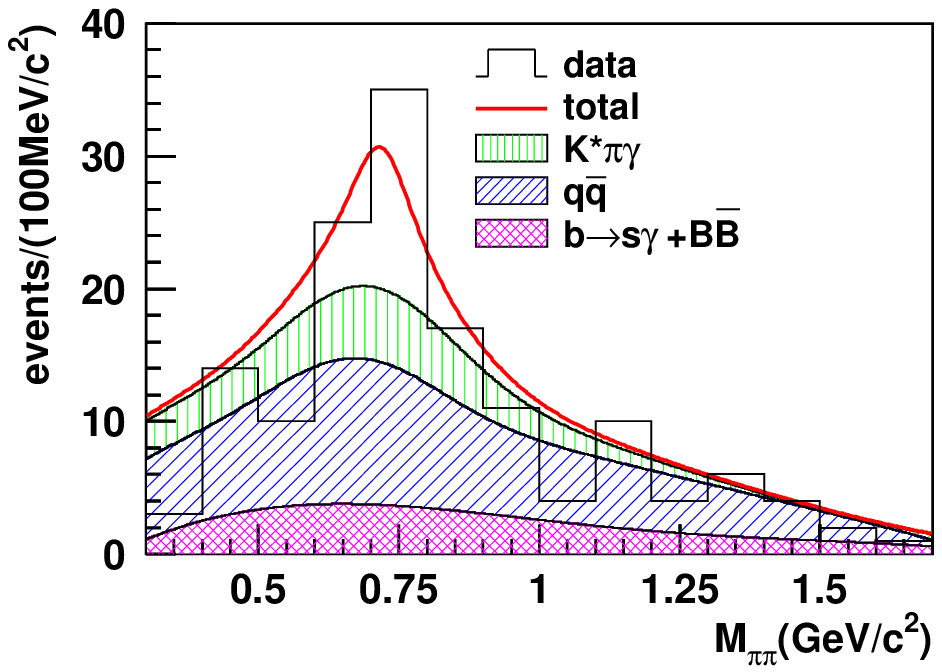}\hss}
\caption{$M(K\pi)$ and $\cos\theta_{\rm hel}$ for
         $B\to K_2^*\gamma$ (upper plots); $M(K\pi)$ and $M(\pi\pi)$ for
         $B\to K\pi\pi\gamma$ (lower plots).}
\label{fig:kpipigam}
\end{figure}

Belle has also explored other exclusive $\btosgamma$ decays as shown in
Fig.~\ref{fig:kpipigam} using a $29\fbinv$ dataset
\cite{bib:belle-kxgam-prl}.  In the $\KP\PM\gamma$ final state, we
clearly observe a tensor component of $B^0\to K_2^{*}(1430)^0\gamma$.
When we add one more pion, we also observe the $B$ decay signal which is
consistent with a sum of $K^*\pi\gamma$ and $K\rho\gamma$ final states.
We obtained branching fractions
\begin{equation}
\begin{array}{l}
{\displaystyle
  \Br(B^0\to K_2^*(1430)^0\gamma)=(13\,\pm5\,\pm1)\times10^{-6}}\\
{\displaystyle
  \Br(B^+\to K^+\pi^+\pi^-\gamma)=(24\,\pm5\,^{+4}_{-2})\times10^{-6}}.\\
\end{array}
\end{equation}
Assuming isospin symmetry to extrapolate to the other charge
combinations, the $K\pi\gamma$ and $K\pi\pi\gamma$ final states cover
about $(35\pm8)\%$ of the total $\BtoXsgamma$ branching fractions, and
hence it is important to understand these decays for a further precision
measurement of inclusive $\BtoXsgamma$.

 % ---------------------------------------------------------

                    %%%%%%%%%%%%%%%%%%%%%%%%%%%%%%%%
                    \section{$\btodgamma$ processes}
                    %%%%%%%%%%%%%%%%%%%%%%%%%%%%%%%%

The $\btodgamma$ process is expected to be observed in one of the
exclusive modes, $B\to\rho\gamma$ and $B\to\omega\gamma$.  The analysis
is similar to the study of $\BtoKstarG$.  Since the expected branching
fraction is almost two orders of magnitude smaller than $\BtoKstarG$,
continuum background is very high.  $\BtoKstarG$ is also a
significant background, since the particle identification, which is
mainly based on the threshold aerogel Cherenkov detector, has almost
10\% kaon to pion fake rate.  Candidate $\rho$
($\omega$) mesons are selected in a window of $\pm150$ $(30)\MeVcc$ around the
nominal mass.  About a half of the $K^*$ with a pion mass hypothesis for
the kaon falls into this $\rho$ mass window.  The $K^*\gamma$ background
is much more severe in the neutral mode, since the sub-decay branching
fraction of $K^{*0}\to\KP\PM$ is twice as high as that of
$K^{*+}\to\KP\PZ$ and the expected branching fraction for
$B^0\to\rho^0\gamma$ is twice as low as that for $B^+\to\rho^+\gamma$.
Therefore, we explicitly reject the events if the $\KP\PM$ invariant mass
(with a kaon mass hypothesis for one of the pions) is in a range of
$\pm50\MeVcc$ of the $K^*$ mass.

\begin{figure} % -------------------------------------------------------
\hbox to\hsize{%
  \hss\includegraphics[width=\hsize]{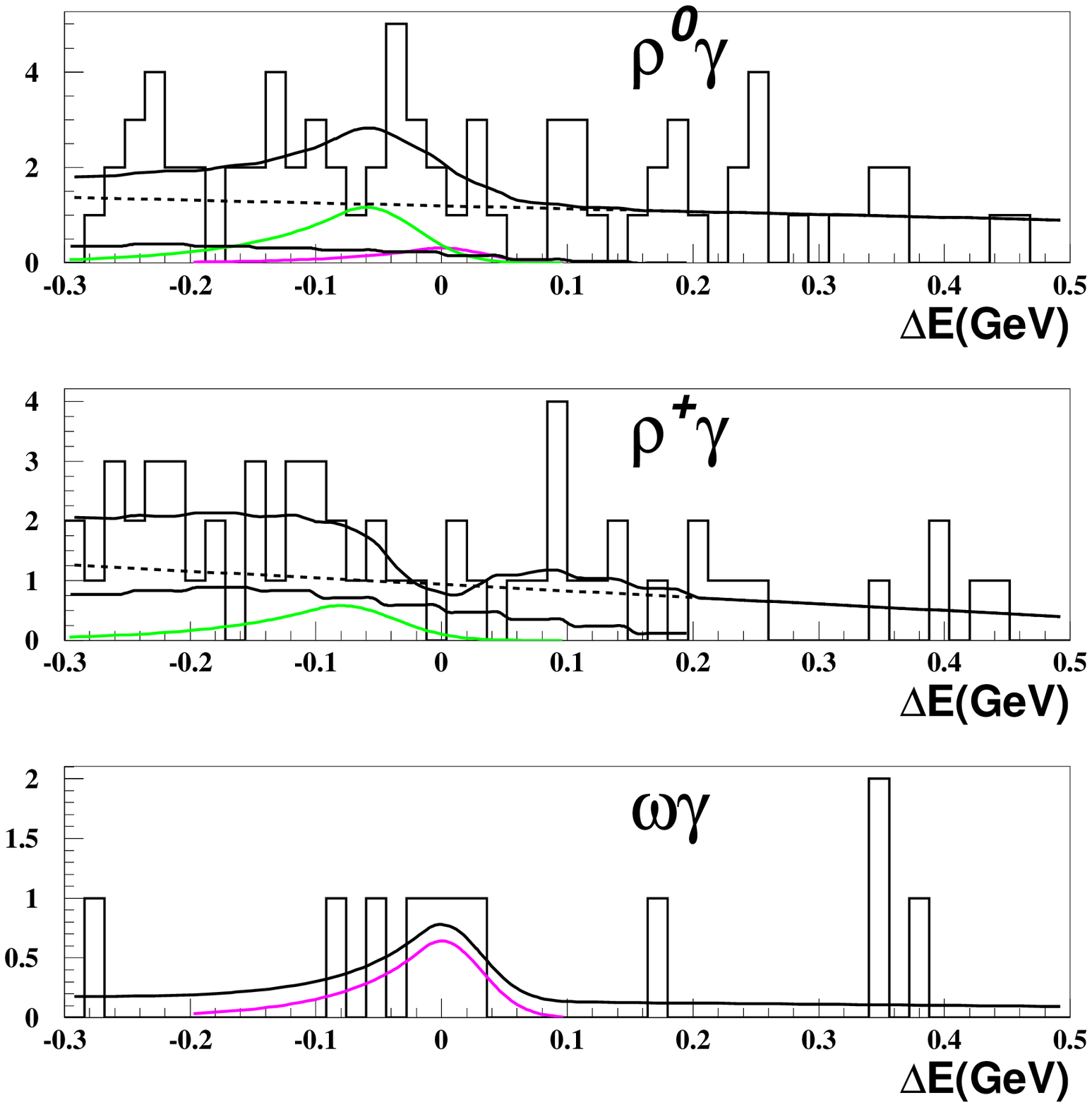}\hss}
\caption{$\DeltaE$ distributions for $B^0\to\rho^0\gamma$,
          $B^+\to\rho^+\gamma$ and $B^0\to\omega\gamma$.}
\label{fig:rhogam}
\end{figure}

Using a $78\fbinv$ dataset, we performed a simultaneous fit to the three
modes of $\rho^+\gamma$, $\rho^0\gamma$ and $\omega\gamma$ (we denote as
$(\rho+\omega)\gamma$) with an assumption of the isospin relation
$\Gamma((\rho+\omega)\gamma)=\Gamma(\rho^+\gamma)=2\Gamma(\rho^0\gamma)=2\Gamma(\omega\gamma)$.
In this case, we neglect isospin violating effects such as the
annihilation diagram contribution that only appears in
$B\to\rho^+\gamma$.  We also perform a simultaneous fit to the three
$(\rho+\omega)\gamma$ modes and two $K^*\gamma$ modes to evaluate the
ratio $\Gamma(B\to(\rho+\omega)\gamma)/\Gamma(B\to K^*\gamma)$. For this
updated analysis, we reoptimized the signal yield extraction method from
the previous analysis with $60\fbinv$ \cite{bib:belle-rhogam-ichep02}.
We decided to use a fit to the distribution of the energy difference
($\DeltaE$), from a toy Monte Carlo study to optimize the result of the
ratio.  After selecting the events in the $2\sigma$ window of $\Mbc$ and
optimizing the continuum suppression cut, which is based on a likelihood
ratio of the $B$ meson flight direction and a Fisher discriminant formed
from a modified set of Fox-Wolfram moments, we fit the $\DeltaE$
distribution with a linear continuum background component, a MC
determined $B$ decay background component and two Crystal Ball line
shapes to represent the signal and the $K^*\gamma$ component (not in the
$\omega\gamma$ mode).  The $\DeltaE$ peak is shifted by $-60\MeV$ for
the $K^*\gamma$ background.  The individual fit result for each mode is
shown in Fig.~\ref{fig:rhogam}.

We observe no significant signal yield in the individual fits nor in the
simultaneous fits, and obtain the following 90\% confidence level upper
limits
\begin{equation}
\begin{array}{l}
{\displaystyle
  \Br(B^+\to \rho^+\gamma)<2.7\times10^{-6}}\\
{\displaystyle
  \Br(B^0\to \rho^0\gamma)<2.6\times10^{-6}}\\
{\displaystyle
  \Br(B^0\to \omega\gamma)<4.4\times10^{-6}}\\
{\displaystyle
  \Br(B\to (\rho+\omega)\gamma)<3.0\times10^{-6}}\\
{\displaystyle
  \Gamma(B\to (\rho+\omega)\gamma)/\Gamma(B\to K^*\gamma)<0.081.}\\
\end{array}
\end{equation}
These results are still a few times larger than the SM predictions, and
the ratio does not give a useful constraint on $|\Vtd/\Vts|$ yet.

 % ---------------------------------------------------------

\begin{figure} % -------------------------------------------------------
\hbox to\hsize{%
  \hss\includegraphics[width=\hsize]{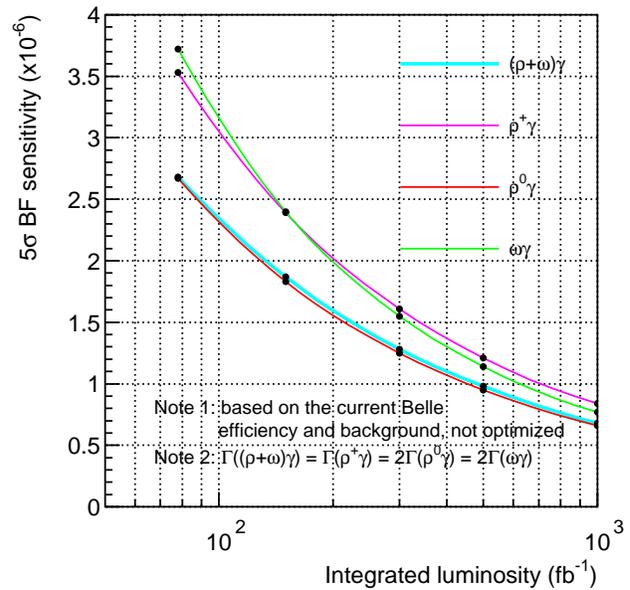}\hss}
\caption{Extrapolated sensitivity for the $5\sigma$ observation of the 
         $B\to(\rho+\omega)\gamma$ signal as a function of the data size.}
\label{fig:rhogam-prospects}
\end{figure}

From the current analysis, one can estimate how much sensitivity we
expect on the search for the $B\to(\rho+\omega)\gamma$ signal.  In
Fig.~\ref{fig:rhogam-prospects}, we use the efficiency and the
size of the backgrounds in the current analysis to extrapolate the
sensitivity for the first observation of the $B\to(\rho+\omega)\gamma$
signal.  It is seen that we need $250$ to $500\fbinv$ of data if the
branching fraction is $1$ to $1.5\times10^{-6}$ as predicted.

Once the first branching fraction measurement is made, there will be a
20 to 30\% experimental error. One has to assume a rather large theory
error on the form factor ratio between $B\to(\rho+\omega)$ and $B\to
K^*$ to interpret the result in terms of $|\Vtd/\Vts|$.  At this moment,
there is no well defined idea to shrink the theory error.

Another possibility is to aim for an inclusive measurement of
theoretically cleaner $B\to X_d\gamma$.  However, no clear idea is known
how one can control the enormous $B\to X_s\gamma$ backgrounds, and this
may not be a possible project in the coming few years.

 % ---------------------------------------------------------

                    %%%%%%%%%%%%%%%%%%%%
                    \section{Conclusion}
                    %%%%%%%%%%%%%%%%%%%%

The long awaited first inclusive measurement of $\BtoXsll$ was performed
by Belle.  Although the results are consistent with the SM, the error is
still large.  The results also demonstrate that future programs
such as the measurement of the forward backward asymmetry are feasible
by adding more data.

Exclusive measurements of $\btosgamma$ processes have been extensively
carried out by Belle.  These results become important to understand the
properties of the hadron recoil system in the next $\BtoXsgamma$
measurement, in which we expect a better understanding of systematic
errors.

The $\btodgamma$ process has not been measured yet, but with the dataset
of about $500\fbinv$, anticipated in the next few years, it is likely to
have the first measurement of the exclusive process
$B\to(\rho+\omega)\gamma$.  This will be important to constrain the
value of $|\Vtd/\Vts|$ without relying on $B^0_{d,s}$-$\Bbar^0_{d,s}$
mixing and lattice calculations.

                    %%%%%%%%%%%%%%%%%%%%%%%%%%
                    \section{Acknowledgements}
                    %%%%%%%%%%%%%%%%%%%%%%%%%%

I would like to express my thanks to the organizers of the Workshop on
the CKM Unitarity Triangle.

\end{document}